\documentclass[preprint,12pt]{elsarticle}

\usepackage{graphicx}
\usepackage{amssymb}
\usepackage{amsmath}

\biboptions{sort&compress}

\journal{Journal of Theoretical Biology}

\begin{document}

\begin{frontmatter}

\title{Oscillation arrest in the mouse somitogenesis clock presumably takes place via an infinite period bifurcation}

\author[Bimed]{Eder Zavala-L\'{o}pez}
\author[Mty,Cambam]{Mois\'{e}s Santill\'{a}n}

\address[Bimed]{Centro de Investigaci\'{o}n y de Estudios Avanzados del IPN, Depto. de Biomedicina Molecular, Av. IPN No. 2508, Col. San Pedro Zacatenco, M\'{e}xico DF, M\'{e}xico}

\address[Mty]{Centro de Investigaci\'{o}n y de Estudios Avanzados del IPN, Unidad Monterrey, V\'{\i}a del Conocimiento 201, Parque PIIT, 66600 Apodaca NL, M\'{e}xico}

\address[Cambam]{Centre for Applied Mathematics in Bioscience and Medicine, 3655 Promenade Sir William Osler, McIntyre Medical Building, Room 1123A, Montreal, QC H3G 1Y6, CANADA}

\begin{abstract}
In this work we address the question of how oscillations are arrested in the mouse somitogenesis clock when the determination front reaches presomitic cells. Based upon available experimental evidence we hypothesize that the mechanism underlying such a phenomenon involves the interaction between a limit cycle (originated by a delayed negative feedback loop) and a bistable switch (originated by a positive feedback loop). With this hypothesis in mind we construct the simplest possible model comprising both negative and positive feedback loops and show that (with a suitable choice of paremeters): 1) it can show an oscillatory behavior, 2) oscillations are arrested via an infinite-period bifurcation whenever the different gene-expression regulator-inputs act together in an additive rather than in a multiplicative fashion, and 3) this mechanism for oscillation arrest is compatible whit plentiful experimental observations.
\end{abstract}


\end{frontmatter}



\section{Introduction}
\label{Intro}

As elegantly reviewed by \citet{Campanelli:2010fk}, somitogenesis is the process by which vertebrate embryos develop somites, which are transient, repeated blocks of cells arising from the presomitic mesoderm (PSM) that differentiate further into vertebrae, ribs, musculature, and dorsal dermis. The tail bud is a proliferative zone at the posterior end of the embryo where immature cells are continually added to the posterior-most PSM. As the tail bud grows away posteriorly, the oldest cells in the anterior PSM segment in groups to form lateral pairs of somites along the midline. The process stops when the anterior formation of somites has progressed posteriorly across the entire PSM (reaching the arresting growth in the tail bud) \citep{Pourquie01,Saga:2001uq,Dequeant:2008fk}.

The first model attempting to explain somitogenesis was due to \citet{CookeZeeman76}. They postulated that the susceptibility of cells in the PSM to form somites oscillates between susceptible and unsusceptible, while a determination wavefront sweeps posteriorly across the PSM. The passing wavefront triggers cells to form somites, but does so only when cells are susceptible, i.e., when their clocks are in the correct phase of oscillation. Since adjacent cells are in phase, cohorts of cells are recruited in succession to form somites. Initially, the clock was thought to be closely linked to the cell cycle \citep{Primmett:1989vn}. However, \citet{PalmerimEtAl97} discovered a gene with oscillatory expression in the PSM of the chick embryo, providing an alternative candidate for the clock. Experimental work has since identified multiple oscillatory genes in each of several model organisms, including mouse \citep{Dequeant:2006ys} and zebrafish \citep{Holley:2007zr}.

In all of these organisms, the oscillatory gene expression in individual cells is coordinated throughout the PSM in order to produce spatiotemporal waves of mRNA and protein expression \citep{DaleEtAl03,Bessho:2003fk}, which we call the clock-wave \citep{PalmerimEtAl97}. Synchronized, periodic expression is observed in the tail bud with a frequency that matches the anterior formation of somites \citep{Dequeant:2008fk,Holley:2007zr}. Broad waves of expression repeatedly initiate in the posterior-most PSM and narrow while traveling anteriorly \citep{Dequeant:2008fk,Saga:2001uq,PalmerimEtAl97}. The waves slow considerably as they reach the region of forming somites. Successive waves arriving at the anterior-most PSM help sequentially establish stable bands of high-low gene expression in several additional genes, indicating nascent somite boundaries and polarity \citep{Dequeant:2008fk,Saga:2001uq}.

Separate experiments have identified biochemical candidates for the wavefront \citep{AulehlaEtAl03, Dubrulle:2004ly, Cinquin:2007uq, GoldbeterEtAl07, Aulehla:2008ve, Aulehla:2010kx}. These bio-molecules exhibit graded concentration profiles across the PSM that shift posteriorly in synchrony with tail bud growth. A changing gradient level triggers mesodermal cell differentiation and somite formation \citep{Dequeant:2008fk,Holley:2007zr,Aulehla:2008ve,AulehlaHerrmann04}. We call this the gradient-wavefront.

The precise mechanism in which the clock-wave interacts with the gradient-wavefront, as well as their possible interactions with intercellular signaling mechanisms, remains unknown \citep{Dequeant:2008fk, Holley:2007zr, AulehlaHerrmann04, GiudicelliLewis04}. Many mathematical models of the dynamics of somitogenesis have been proposed, with reviews and comparisons of several prominent models available in the literature \citep{GoldbeterEtAl07, AulehlaHerrmann04, Dale:2000dq, McInerney:2004vn, GiudicelliLewis04, Kulesa:2007bh, RodriguezGonzalesEtAl07, Santillan:2008nx, Baker:2008qf, Mazzitello:2008ys, Baker:2009cr}.

Previous works have suggested the presence of bistable switches arising from positive feedback loops (formed by mutual interactions between genes involved in the somitogenesis process) \citep{GoldbeterEtAl07, Aulehla:2010kx}. In a recent work \citep{Santillan:2008nx} we suggested that the interaction between an oscillatory regime (originated by a time-delayed negative-feedback loop) and a bistable domain (originated by a positive feedback loop involving genes under different pathways) may be responsible for the arrest of oscillations in presomitic cells in such a way that cohorts of presomitic cells stop oscillating simultaneously. Nevertheless, such a model falls short to explain the experimentally-observed oscillation-period increase as PSM cells approach the somite-formation region \citep{PalmerimEtAl97, Giudicelli:2007zr, Gibb:2009fk}.

In this paper we develop a biologically informed, yet minimally constructed, mathematical model which nonetheless is capable of explaining most of the observed dynamic features of oscillation arrest, and provides a plausible mechanism for this phenomenon; namely, an infinite period bifurcation.

\section{Methods}
\label{meth}

\subsection{Model Development}

Several studies on mice have demonstrated that the expression of various genes under the Notch regulatory pathway oscillates in presomitic cells \citep{Bessho:2003fk, Aulehla:1999qf, Jouve:2000bh, IshikawaEtAl04}. Other modeling studies have suggest that the underlying mechanism is a simple negative feedback loop with relatively long time delays due to transcription and translation of the corresponding genes \citep{Aulehla:2008ve,AulehlaHerrmann04,GiudicelliLewis04,Baker:2008qf}. On the other hand, the expression of genes under the Wnt and FGF signaling pathways has also been experimentally demonstrated to oscillate \citep{Dequeant:2006ys, AulehlaEtAl03, Dale:2006dq, Wahl:2007cr, Geetha-Loganathan:2008nx}, with the same period as that of genes under the Notch pathway. There is evidence that such oscillation synchronization is due to the mutual regulation of genes under the different regulatory pathways \citep{Masamizu:2006tg, Niwa:2007oq, Goldbeter:2008kl}. The most complete regulatory regulatory network involving genes known to participate in the somitogenesis process in the mouse is reviewed in \citep{Dequeant:2008fk}.

Based on the above discussed facts we proposed in a previous work \citep{Santillan:2008nx} a genetic circuit composed of a delayed negative feedback regulatory loop, capable of generating an oscillatory behavior, and a positive feedback loop, behaving as a bistable switch. We observed that the interaction of the switch and the oscillator can explain the simultaneous arrest of oscillations in cohorts of cells; a necessary condition for explaining somite formation. Nevertheless, the model failed to explain the experimentally-observed increase of the oscillation period as presomitic cells get away from the tail bud \citep{PalmerimEtAl97, Giudicelli:2007zr, Gibb:2009fk}. Under the consideration that such period increase is more than just an interesting dynamic feature, we decided to investigate its origin in the present work. 

The simplest possible gene network comprising a negative and a positive feedback loops is that illustrated in Figure \ref{SchemeResults}A. Notice that we have included an open-loop regulatory input, $k$, to account for the wavefront signal. In the previous paper we assumed that a negative and a positive regulatory signals acting upon a given gene only allow its expression when the first is absent and the second is present. In the present work we consider that the gene can be expressed when either the negative signal is absent or the positive signal is present. Given the above considerations, the equation governing the dynamics of the gene circuit in Figure \ref{SchemeResults}A is:
\begin{equation}
\frac{dx}{dt} =  a \left(k+\left( b-\frac{k}{2} \right) f\left( x_{\tau_{n}}\right) +\left( 1-b-\frac{k}{2} \right)g\left( x_{\tau_{p}}\right)-x\right).
\label{modeq}
\end{equation}
In the above equation $f(x)\leq 1$ and $g(x)\leq 1$ respectively stand for the negative and positive feedback loops, $x_{\tau}$ denotes variable $x$ delayed a time $\tau$, $a$ is the degradation rate for proteins $x$, $b$ is the strength of the negative loop as compared to the positive loop, and $0 \leq k \leq 1$ is the strength of the wavefront signal. Equation (\ref{modeq}) has been parametrized in such a way that all of the $x$ input regulatory signals add up to one when they reach their maximal values. This parametrization guaranties that $x$ is a dimensionless variable, and that $x\leq 1$ at all times. To ensure that the sum of all regulatory inputs is always less than one, we have further assumed that the strength of both the negative and the positive loops decreases proportionally to $k$; indeed, we have supposed that both of them decrease in the same amount. Finally, we have assumed that functions $f(x)$ and $g(x)$ are Hill-type functions of the form:
\begin{eqnarray}
f(x) & = & \frac{k_{1}^{n}}{k_{1}^{n}+x^{n}},  \label{fx}\\
g(x) & = & \frac{x^{n}}{k_{2}^{n}+x^{n}}. \label{gx}
\end{eqnarray}
Observe that $f(x)$ is a monotonously decreasing function while $g(x)$ is a monotonously increasing function, in accordance with the fact that they stand for negative and positive regulation, respectively. 

\subsection{Numerical Methods}

All the solutions and numerical analyses performed on the mathematical model here presented were carried out using the software \texttt{xppaut} and the numerical methods available in it \citep{Ermentrout:2002ly}.

\section{Results}

\label{res}
\begin{figure}[htb]
\begin{center}
\includegraphics[width=3in]{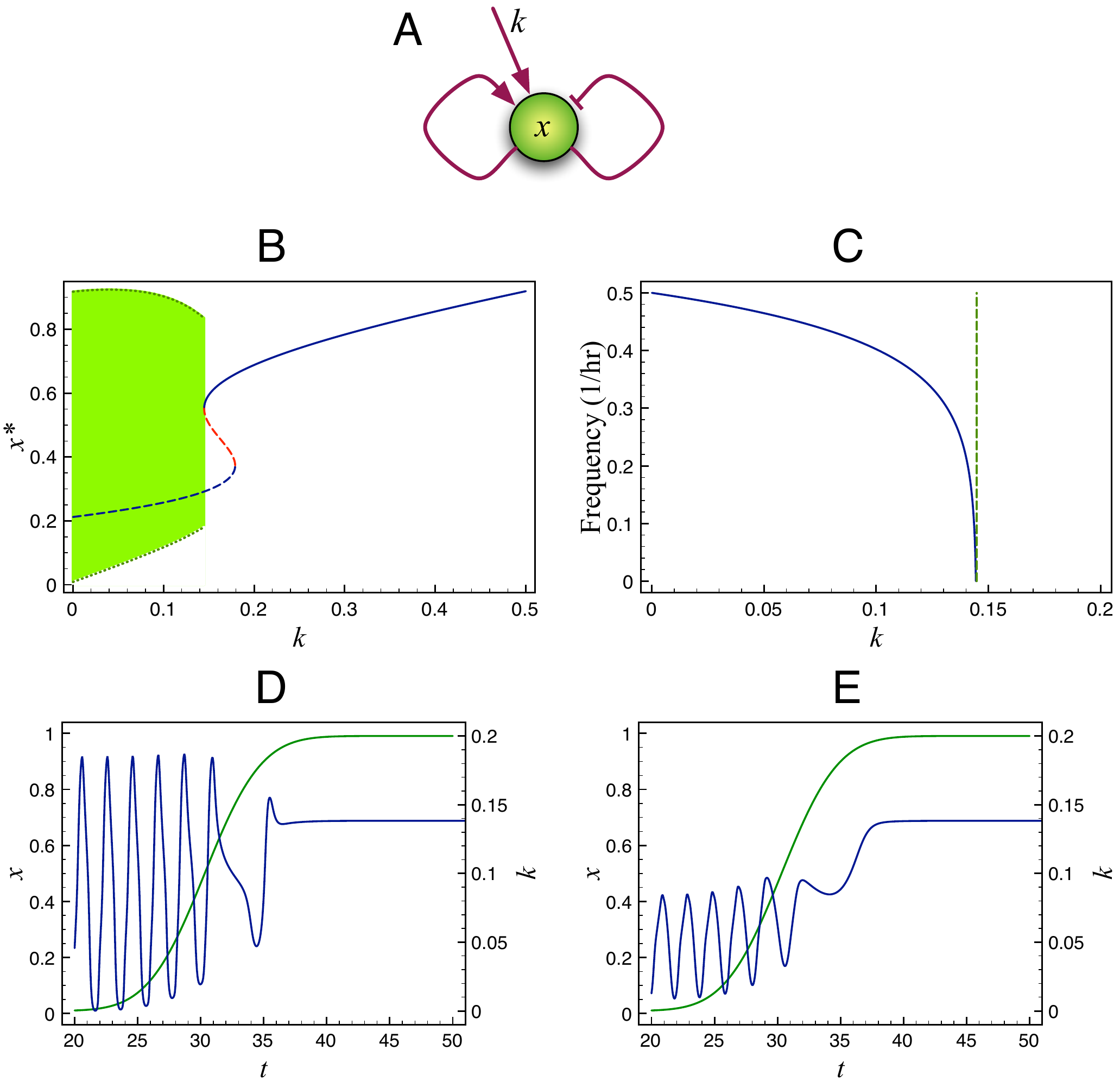}
\end{center}
\caption{A. Schematic representation of the gene network studied in this work. B. Bifurcation diagram of the mathematical model given by Eqns. (\ref{modeq})-(\ref{gx}), with the model parameters set in such a way that an infinite period bifurcation is obtained; in particular $\tau_{p}=0$. C. Plot of the oscillation frequency vs. the bifurcation parameter $k$, with the same parameter values as in B. D. Plot of $x$ vs. $t$ (blue line) calculated with the same parameter values as in C, and with $k(t)$ as plotted (green line). E. Same as in D, but with $\tau_{p}=20\, \text{min}$.}
\label{SchemeResults}
\end{figure}

Of all the parameters in the model we fix $n=4$, which corresponds to a biologically plausible level of cooperativity, and analyze the model dynamics under variations of the rest of them. We start by studying the behavior of the negative feedback loop by itself, and thus we set $b=1$ and $k=0$. As previously reported, we found an oscillatory behavior for a vast set of $a$, $k_{1}$ and $\tau_{n}$ values. Moreover, the smaller the value of $\tau_{n}$, the larger the value of $a$ and the smaller the value of $k_{1}$, in order for the system to present sustained oscillations. Finally, $\tau_{n}$ is the parameter that has, by far, the largest influence on the oscillation period. Thus, we selected $\tau_{n}=0.64 \, \text{hrs} = 38.4 \, \text{min}$, $a=10 \, \text{hrs}^{-1}$, and $k_{1}=0.24$ to reproduce the $2\,\text{hrs}$ oscillation period reported for the mouse somitogenesis clock.

\begin{figure}[htb]
\begin{center}
\includegraphics[width=1.5in]{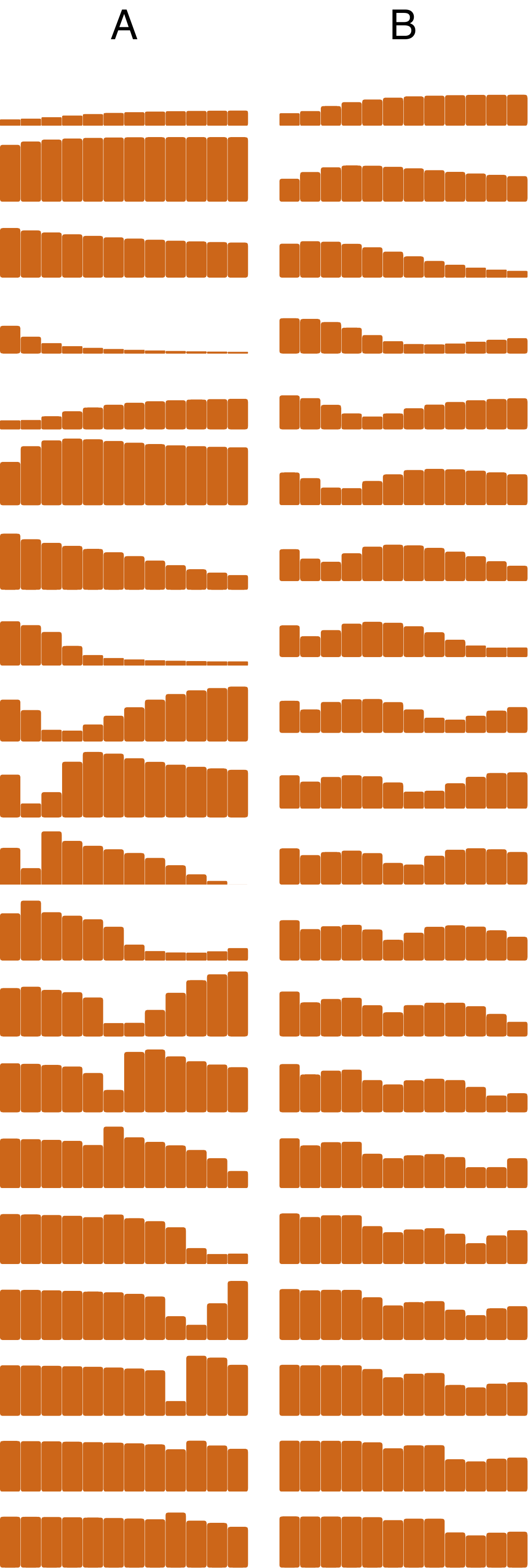}
\end{center}
\caption{Snapshots, taken every $30 \, \text{min}$, of the time evolution of 12 cells aligned along the PSM, with a separation of $30 \, \text{min}$ between consecutive cells. The time span of the whole simulation begins when all the cells are in the tail bud and finishes when the last cell is arresting its oscillations. The simulation in A was computed with $\tau_{p}=0$ (the height of each column is proportional to the value of $x$ in the corresponding cell), while the simulation in B was carried out considering $\tau_{p}=20 \, \text{min}$. }
\label{SimuImgs}
\end{figure}

To analyze the interaction between the oscillator and the bistable switch we study the model behavior under variations of parameters $k$ and $b$. Parameter $k$ plays the role of a bifurcation parameter since the oscillations of $x$ are invariably arrested as the value of $k$ is increased beyond a given threshold. In many cases, oscillation arrest occurs via a Hopf-like bifurcation. That is, the oscillation amplitude decreases continuously as $k$ increases, until it eventually becomes zero. However, for some values of parameters $b$ and $k_{2}$ the oscillation arrest occurs via an infinite period bifurcation. After thoughtfully browsing the parameter space we found that parameter $b$ needs to take values in the range $[0.2,0.4]$ in order to have an infinite period bifurcation. For values larger than $0.4$, the oscillations are arrested via a Hopf-like bifurcation, while for $b<0.4$ no oscillatory behavior is observed whatsoever. Regarding $k_{2}$ the range of values this parameter can attain so we observe an infinite period bifurcation depends on the value of $b$. For instance, when $b=0.2$ an infinite period bifurcation occurs whenever $k_{2}\in[0.46,0.49]$, when $b=0.3$ the range of $k_{2}$ values for which an infinite period bifurcation is observed is $[0.41,0.52]$, and when $b=0.4$ the condition for having an infinite period bifurcation is that $k_{2}\in[0.37,0.43]$. In all cases, the $k$ bifurcation value is a growing function of $k_{2}$. In Figure \ref{SchemeResults}B we present the bifurcation diagram calculated with the following $b$ and $k_{2}$ parameter values: $b=0.3$ and $k_{2}=0.48$; while in Figure \ref{SchemeResults}C we plot the oscillation frequency as a function of $k$, for the same parameter values. In this last plot we can see how the frequency goes to zero (and so the period goes to infinity) as $k$ reaches its bifurcation value.

As the embryo grows the tail bud recedes and, while doing so, leaves some cells behind. After a cell leaves the tail bud, the levels of the different morphogenic substances (Fgf8, Wnt3a, and Retinoic Acid) start changing: the Fgf8 and Wnt3a levels decrease,  while the Retinoic Acid level increases \citep{Aulehla:2010kx}. In our model we account for those changes by increasing the value of parameter $k$. In Figure \ref{SchemeResults}D we illustrate the behavior of an oscillatory presomitic cell as the value of parameter $k$ increases nonlinearly with time; starting from zero and up to a value 4/3 times its bifurcation value. Notice how the oscillation period increases together with $k$, and how, when $k$ surpasses its bifurcation value, the system completes one further oscillation and then reaches a steady state.

To better understand the process of oscillation arrest along the PSM we simulate, following \citep{Santillan:2008nx}, the time evolution of 12 cells aligned along the PSM, with a separation of 30 min between consecutive cells. If the tail bud recedes at constant velocity and keeps leaving cell behind in a steady fashion, the distance between two PSM cells is proportional to the difference of elapsed times since their leaving the TB. Given that the oscillation period is 2 hrs, the considered cell set spans a PSM region three periods long. To account for the cell separation, we simply assumed in our simulations that the function describing how parameter $k$ decays in time is delayed in proportion to how much later a given cell left the TB. In other words, if $k(t)$ describes the time evolution of parameter $k$ for the first cell, $k(t-(i-1)\Delta T)$ is the corresponding function for the $i$-th cell, with $\Delta T = 30\,  \text{min}$.

The results of the simulations described in the previous paragraph are plotted in Figure \ref{SimuImgs}A. There, we present from top to bottom snapshots taken every 30 min of the state of all 12 cells. In total, these snapshots account for five cycles. Observe that at first all the cells oscillate synchronously. This happens because $k\simeq 0$ for all of them, simulating their stay in the tail bud. However, as time passes and the frontmost cells leave the tail bud, their oscillation period decreases until they stop oscillating and the gene expression remains at a high level. Another feature worth noticing is the fact that the oscillation arrest takes place in cohorts of four cells. That is, the first four cells complete three cycles before stopping, while the second and third four-cell sets complete four and five oscillations, respectively. The process described above can be more clearly visualized in the movie Mov1 of the Supplementary Material.

The origin of the time delay $\tau_{n}$ has been attributed to the sum of the times taken by transcription, mRNA splicing, mRNA translocation, translation, post-transductional modification, diffusion of transcription factors into the nucleus, etc. Since all of these processes also take place in the positive feedback regulation loop, a time delay must be also associated to it. Let us denote such a time delay by $\tau_{p}$. We investigated the influence of parameter $\tau_{p}$ on the system behavior and found that it is qualitatively the same whenever $\tau_{p} \leq 20 \, \text{min}$. In particular, the cyclic behavior disappears as $k$ is increased beyond a given value, via an infinite period bifurcation. Larger $\tau_{p}$ values render a more complex behavior whose analysis is beyond the scope of the present paper.

We calculated the behavior of the model with $\tau_{p}=12\, \text{min}$ when $k$ increases nonlinearly with time; starting from zero and up to $0.2$. This simulates a single presomitic cell from the time it is in the tail bud, until it stops oscillating. The results are plotted In Figure \ref{SchemeResults}E. Observe that the results are qualitatively equivalent to those in Figure \ref{SchemeResults}D. The most noticeable differences are that the oscillation amplitude is smaller when $\tau_{p}>0$, and that the gene expression level increases after the oscillations are arrested.

To test the influence of the time delay $\tau_{p}$ on the process of oscillation arrest along the PSM, we repeated the simulations in Figure \ref{SimuImgs}A, but now with  $\tau_{p}=12\, \text{min}$. The results are plotted in Figure \ref{SimuImgs}B. A comparison of Figures \ref{SimuImgs}A and \ref{SimuImgs}B reveals that the behavior of both models is qualitatively equivalent. The only noticeable differences are that, in the model accounting for a time delay in the positive feedback loop, the oscillation amplitude is smaller, the increase of the gene expression level after the oscillations are arrested is more notorious, and the cell cohorts are more clearly defined when they stop cycling. The process described above can be more clearly visualized in the movie Mov2 of the Supplementary Material.

\section{Conclusions}
\label{conclu}

Under the assumption that oscillation arrest in the somitigenesis clock of the mouse takes place due to the interaction between the oscillator and a bistable switch, we have constructed and analyzed the simplest possible model that accounts for this interaction. Although the resulting model is so simple that no quantitative predictions can be expected, we are convinced that it captures enough detail of the interactions between some of the genes involved in the somitogenesis processes to render general qualitative predictions.

The results of the model agree with the following experimental observations regarding oscillation arrest in presomitic cells:
\begin{enumerate}
\item All cells oscillate in synchrony while they are in the tail bud, but their oscillation period increases with time after they leave the tail bud and enter the presomitic mesoderm.

\item PSM cells stop oscillating in equally sized cohorts. These cohorts in turn give rise to somites.

\item The expression levels of the genes under the Notch pathway increases after they stop oscillating, and they reach a stationary state of high expression \citep{IshikawaEtAl04, AulehlaEtAl03, Yasuhiko:2006ve}.
\end{enumerate}

To our consideration, the above enumerated results validate the model to a level that makes feasible the derivation of biological conclusions out from it. In particular, our model results suggest that oscillation arrest takes place via an infinite period bifurcation, that a positive and a negative feedback loop need to act together (in an additive, rather than in a multiplicative fashion) upon a single gene in order for this bifurcation to occur. The model predictions regarding the existence of interacting positive and negative feedback loops, and that regarding the additivity of both loops can be experimentally tested in principle.


\bibliographystyle{model1-num-names}
\bibliography{SnicSomit}

\end{document}